\documentclass[preprint,sort&compress,12pt]{elsarticle}

\usepackage{setspace}
\usepackage{amsmath}
\usepackage{amssymb}
\usepackage{bm}
\usepackage[ruled,vlined]{algorithm2e}
\usepackage{algorithmic}
\usepackage[margin=1in]{geometry}
\usepackage{xcolor}
\usepackage[caption=false]{subfig}
\usepackage{multirow}

\usepackage{lineno}
\usepackage{todonotes}
\biboptions{comma,square}
\usepackage{graphicx}
\usepackage{subfig}
\usepackage{floatrow}
\usepackage{diagbox}
\usepackage{lineno}

\usepackage{caption}
\journal{arXiv}
\newcommand{\norm}[1]{\left\lVert#1\right\rVert}
\begin{document}

\begin{frontmatter}



\title{Physics-Constrained Bayesian Neural Network for Fluid Flow Reconstruction with Sparse and Noisy Data}

\author{Luning Sun}
\author{Jian-Xun Wang\corref{corxh}}

\address[ndAME]{Department of Aerospace and Mechanical Engineering, University of Notre Dame, Notre Dame, IN}
\address[ndCICS]{Center for Informatics and Computational Science, University of Notre Dame, Notre Dame, IN}

\cortext[corxh]{Corresponding author. Tel: +1 574-631-5302}
\ead{jwang33@nd.edu}

\begin{abstract}
In many applications, flow measurements are usually sparse and possibly noisy. The reconstruction of a high-resolution flow field from limited and imperfect flow information is significant yet challenging. In this work, we propose an innovative physics-constrained Bayesian deep learning approach to reconstruct flow fields from sparse, noisy velocity data, where equation-based constraints are imposed through the likelihood function and uncertainty of the reconstructed flow can be estimated. Specifically, a Bayesian deep neural network is trained on sparse measurement data to capture the flow field. In the meantime, the violation of physical laws will be penalized on a large number of spatiotemporal points where measurements are not available. A non-parametric variational inference approach is applied to enable efficient physics-constrained Bayesian learning. Several test cases on idealized vascular flows with synthetic measurement data are studied to demonstrate the merit of the proposed method. 
\end{abstract}


\end{frontmatter}


\section{Introduction}
\label{introduction}
Reconstruction of a flow field from limited and noisy measurements is of great significance yet challenging in many engineering applications. For example, the rapid development in flow magnetic resonance (MR) imaging techniques enables noninvasive assessment of hemodynamic information for cardiovascular research and healthcare~\cite{lawley20184d}. However, the resolution and signal-to-noise ratio (SNR) of MR images still remain the limiting factors for clinical applications~\cite{callaghan2017spatial}. Similar scenarios can also be found in monitoring wind farms or other aerodynamic systems, where measurement sensors (e.g., lidar) are usually placed at sparse locations and thus the collected data are also sparse and noisy~\cite{krishnamurthy2013coherent}.        

Because of its wide range of applications, full-field reconstruction of sparse, noisy flow data has become an active research area and received a great deal of attention. In order to compensate for the incompleteness and sparsity of the gappy data, additional information is required, which can be obtained either from an offline flow database or a physics-based model. Based on what type of information is incorporated, the existing flow reconstruction (i.e., superresolution) methods can be organized into two groups: (i) When large offline full-field flow data sets are available, the coherent structures and correlation features of the fluid flow can be extracted, which will be utilized to reconstruct the high-resolution flow fields from sparse online data. Proper orthogonal decomposition (POD)~\cite{sirovich1987turbulence} or dynamic mode decomposition (DMD)~\cite{schmid2010dynamic,tu2013dynamic} are commonly used for flow feature extraction. For instance, Gappy POD has been applied for steady and unsteady flow-field reconstruction in various applications~\cite{venturi2004gappy,bui2003proper,podvin2006reconstruction,yakhot2007reconstruction,moreno2016aerodynamic,mifsud2019fusing}. To overcome the linearity limitations of POD and DMD, deep learning based approaches (e.g., autoencoder neural networks) have been recently developed to extract nonlinear latent representations of the flow field from massive offline data~\cite{lee2019model}. As an alternative, sparsity-promoting representation techniques, e.g., compressed sensing, have also been demonstrated to be able to achieve the same goal more robustly when data is noisy~\cite{callaham2019robust,manohar2018data}. All the algorithms described above rely on a large number of flow datasets for offline ``training", which might not be available in many cases. (ii) Instead of learning from the offline database, the other type of flow reconstruction methods takes advantage of a physics-based model, e.g., computational fluid dynamics (CFD) model, which is able to provide full-field flow predictions. The sparse measurement data can be fused into the model-based predictions using data assimilation (DA) techniques, e.g., ensemble Kalman filter, particle filters, or variational based DA techniques~\cite{foures2014data,combes2015particle,kikuchi2015assessment,mons2016reconstruction,symon2017data,wang2016data}. Nonetheless, physics-based simulations are time-consuming in general while the DA process usually involves numerous model evaluations, which could be computationally prohibitive.     

The recent advances of deep learning techniques for image superresolution~\cite{7839189,dong2014learning} open up new avenues for developing efficient algorithms of flow reconstruction from limited measurements. For example, neural networks (NN) have been used to learn POD coefficients~\cite{yu2018flowfield} or directly capture an end-to-end mapping between the sparse measurements and the high-resolution flow field~\cite{erichson2019shallow}. However, the success of these deep learning models is mostly dependent on a sufficient amount of offline training data, which, as mentioned above, are inaccessible in many applications, e.g., superresolution for flow MR imaging. To alleviate data sparsity, a physics-constrained deep learning strategy has been proposed~\cite{raissi2019physics,zhu2019physics,SUN2019112732,gao2019non}, where physical laws of a system (e.g., Navier-Stokes equations in fluid mechanics) are leveraged to constrain the training process. Recently, this idea has attracted increasing attention and its merits have been demonstrated in solving a number of forward and inverse problems governed by classic partial differential equations (PDEs). Notably, the physics-informed neural networks (PINN) proposed by Rassi et al.~\cite{raissi2019physics} were applied to reconstruct a flow field by assimilating scalar concentration data of a flow field~\cite{raissi2018hidden}. Sun et al. developed a PINN-based fluid surrogate model with encoded boundary conditions and demonstrated that the flow solutions of parametric Navier-Stokes equations can be learned without using any labeled training data~\cite{SUN2019112732}. Although the physics-constrained deep learning shows great promise for flow reconstruction of limited data, the measurement noise associated with the data and model-form uncertainties due to model inadequacy cannot be considered since the classic deep learning models are usually formulated in a deterministic way. Researchers have recently started to explore the uncertainty quantification (UQ) analysis of physics-constrained deep learning by using arbitrary polynomial chaos~\cite{zhang2019quantifying} and variational inference~\cite{zhu2019physics,yang2019adversarial,geneva2020modeling}.

In this work, a physics-constrained Bayesian neural network (PC-BNN) is proposed for flow field reconstruction from sparse and noisy measurements. In contrast to previous works, the equation-constrained training is formulated in a Bayesian manner, where the posterior distribution of the NN weights will be obtained based on the likelihood function,  which is defined by the uncertainty from both measurement noise and model inadequacy. Specifically, the confidence of the physical/physiological constraints is modeled in a probabilistic way, being combined with data uncertainty to form the likelihood function~\cite{wu2019adding}. A non-parametric variation inference algorithm, Stein variation gradient decent (SVGD)~\cite{liu2016stein}, is adopted to efficiently perform the Bayesian learning with limited training overhead compared to its deterministic version. The merit of the proposed method is demonstrated on the reconstruction of idealized vascular flows with sparse and noisy velocity data. The rest of the paper is organized as follows. The proposed physics-constrained Bayesian neural network for flowfield reconstruction is introduced in Section~\ref{methodology}. Numerical studies on test flows with two idealized vascular geometries are presented in Section~\ref{result}. The roles of data and physical constraints in deep learning will be discussed in Section~\ref{discussion}. Finally, Section~\ref{conclusion} concludes the paper.

\section{Methodology}
\label{methodology}

\subsection{Overview}
The general idea of this work is to reconstruct a high-resolution flow field from low-resolution (sparse or possibly noisy) measurement data based on deep neural networks (DNN). Instead of training the DNN on extra offline databases of high-resolution flow fields, physical/physiological principles are leveraged to constrain the learning process and provide additional information for super-resolution. Namely, a pointwise DNN model will be trained on sparse velocity data to capture the flow field. In the meantime, the physical laws are imposed on a large number of spatiotemporal collocation points where measurements are not available. Therefore, the trained DNN is a smooth function in spatiotemporal space and can reconstruct the flow field with arbitrarily high resolution. The physics-constrained deep learning is usually formulated as a \emph{deterministic} optimization problem, where a loss function is defined by combining both the data mismatches and the residuals of governing equations of a physical model, e.g., incompressible Navier-Stokes equations for Newtonian flows~\cite{raissi2018hidden, SUN2019112732}. By minimizing the physics-informed loss, the solution is expected to satisfy the physical model as well as match the training data. This formulation here is referred to as the \emph{deterministic physics-constrained deep learning}. However, when the physical model is not perfect and noisy data are used, the prediction uncertainty regarding model inadequacy and measurement noise cannot be considered in such a deterministic learning process. To address this issue, we developed a \emph{probabilistic physics-constrained Bayesian learning} framework, where the physics-constrained training is formulated in a Bayesian way. Instead of defining the loss, a physics-informed \emph{likelihood} function is constructed, where the measurement noise and equation residuals are modeled as random variables with specified distributions. Given the physics-informed likelihood and specified prior information (DNN initialization), the posterior distribution of the DNN weights can be computed based on the Bayes's theorem. Considering the high dimensionality of DNN, variational inference (VI) is employed to enable feasible Bayesian deep learning. All these components are described further below.

\subsection{Deterministic physics-constrained deep learning}
As mentioned above, a DNN approximator $\mathbf{f}^\theta(t, \mathbf{x}) = [\mathbf{u}^\theta, P^\theta]$ is constructed to capture the true pointwise flow solution $\tilde{\mathbf{f}}(t, \mathbf{x}) = [\tilde{\mathbf{u}}, \tilde{P}]$, where $\mathbf{u}, P$ represent velocity and pressure, and $\theta$ represents DNN parameters (e.g., weights and bias). The training of this neural network relies on two pieces of information: sparse (noisy) velocity data $\mathbf{u}^d$ and a physical model of the fluid system. The data-based loss component can be defined straightforwardly as the data mismatch, $\norm{\mathbf{u}^\theta - \mathbf{u}^d}$, while the physics-based loss component is built upon the fluid governing equations. Here, we model the fluid dynamics by a set of incompressible Navier-Stokes equations with the Newtonian assumption,
\begin{equation}
	\label{eq:ns}
	\mathcal{R}(\mathbf{u}, P) = 0 := \left \{
	\begin{aligned}
	&\underbrace{\nabla \cdot \mathbf{u} = 0}_{\mathrm{Mass \ conservation}},  &\mathbf{x}, t \in \Omega_f \times [0, T],\\
	&\underbrace{\frac{\partial\mathbf{u}}{\partial t} + (\mathbf{u}\cdot\nabla)\mathbf{u} + \frac{1}{\rho}\nabla P - \nu\nabla^2\mathbf{u} + \mathbf{b}_f = 0}_{\mathrm{Momentum \ conservation}}, \quad   &\mathbf{x}, t \in \Omega_f \times [0, T], 
	\end{aligned} \right .
\end{equation}
where $\mathcal{R}(\mathbf{u}, P)$ represents the residual function of the Navier-stokes equations; $t$ and $\mathbf{x}$ are temporal and spatial coordinates, respectively; $\rho$ and $\nu$ are density and viscosity of the fluid, respectively; $\mathbf{b}_f$ is the body force; To determine unique flow solutions, proper initial ($\mathcal{I}(\mathbf{u}, P)=0$) and boundary conditions ($\mathcal{B}(\mathbf{u}, P)=0$) are required. The DNN-approximated solutions $[\mathbf{u}^\theta, P^\theta]$ are also expected to comply with the physical model, and thus the violation of the Eq.~(\ref{eq:ns}) will be penalized as well. Hence, the physics-regularized loss function can be defined as,
\begin{equation}
\label{eq:loss}
    \mathcal{L}(\theta) = \norm{\mathcal{R}(\mathbf{u}^\theta, P^\theta)} + \lambda_d\norm{\mathbf{u}_\theta - \mathbf{u}_d}
\end{equation}
where all the derivative terms in $\mathcal{R}$ are computed using automatic differentiation and $\lambda_d$ is a trainable penalty coefficient. The physics-constrained training is defined as a constrained optimization problem,
\begin{equation}
	\begin{aligned}
	&\theta^* = \underset{\theta}{\arg\min} \ \mathcal{L}(\theta),\\
		&\mathrm{s. t.}
		\left \{
		\begin{split}
		&\mathcal{I}(\mathbf{x}, P^\theta, \mathbf{u}^\theta) = 0,  \qquad &t =0, \mathrm{in}\ \Omega_f,\\
		&\mathcal{B}(t, \mathbf{x}, P^\theta, \mathbf{u}^\theta) = 0, &\mathrm{on}\ \partial\Omega_f,
		\end{split} \right.
	\end{aligned}
\end{equation}
To impose the initial and boundary conditions (IC\&BC), two strategies can be used: (i) IC\&BC are formulated as additional penalty terms into the loss function and imposed in a soft manner; or (ii) they can be encoded into the DNN structure in a hard manner as shown in Ref.~\cite{SUN2019112732}. In this work, the pressure inlet/outlet boundary conditions will be enforced by construction while the no-slip wall boundary condition is imposed softly to avoid involving additional networks for complex geometries. In general, the data loss can only be computed on a handful of points due to data sparsity, but the residual of the physical model will be penalized on a large number of points randomly selected from the physical domain. The Adam stochastic gradient descent (SGD) algorithm~\cite{kingma2014adam} is used to solve this optimization problem.

\subsection{Probabilistic physics-constrained Bayesian learning}
The deterministic formulation of physics-constrained DNN has limitations when it comes to noisy data and imperfect physical models. To reflect uncertainties associated with the data and model, a probabilistic formulation should be considered, where the training is conducted in a Bayesian way. Namely, the DNN $\mathbf{f}^\theta(t, \mathbf{x})$ is initialized by specifying a prior distribution $p(\theta)$ for network parameters $\theta$. By constructing the likelihood function $p(\mathcal{D}, \mathcal{R}|\theta)$ based on the sparse data $\mathcal{D} = \{\mathbf{u}^d\}$ and physical model $\mathcal{R}=0$, the posterior distribution $p(\theta|\mathcal{D}, \mathcal{R})$ can be obtained using Bayes' rule,
\begin{equation}
p(\theta|\mathcal{D}, \mathcal{R}) \propto p(\theta)p(\mathcal{D}, \mathcal{R}|\theta).
\end{equation} 
By sampling the posterior, the trained DNN can provide a mean prediction as well as estimated uncertainties.

\subsubsection{Bayesian learning using non-parametric variational inference}
\label{method:VI}
Although efficient Monte Carlo sampling approaches such as Markov Monte Carlo (MCMC) are standard for Bayesian inference and have been widely used to approximate the posterior distribution, they are usually infeasible for an extremely high-dimensional problem like DNN training, which may involve millions of parameters. Variational inference (VI), instead, recasts the Bayesian inference as a deterministic optimization problem by minimizing the Kullback-Leibler (KL) divergence between a proposed distribution $q(\theta)$ and the target distribution (i.e., posterior distribution) as,  
\begin{equation}
\label{eq:1}
\theta^* = \underset{\theta}{\arg\min} \ \mathbb{KL}\left(q(\theta)||p(\theta|\mathcal{D}, \mathcal{R})\right) = \underset{\theta}{\arg\min} \  \mathbb{E}_{q}\left[\log\left(\frac{q(\theta)}{p(\theta|\mathcal{D}, \mathcal{R})}\right)\right],
\end{equation} 
where $\mathbb{E}_q(\cdot)$ is the expectation with probability density $q$. The KL divergence is a measure of the discrepancy between two probability distributions. Most often, the proposed density is parameterized with a specified form of distributions. The performance of the parametric VI largely depends on the predefined family of distributions, which introduces deterministic biases~\cite{Blundell:2015:WUN:3045118.3045290}. In this work, a non-parametric VI method, Stein variation gradient descent (SVGD)~\cite{liu2016stein}, is adopted, which uses a set of $n$ particles $\{\theta_i\}_{i=1}^n$ to directly minimize the KL divergence without the need of defining variational approximation family. The general idea is to iteratively move the set of particles towards the posterior distribution using the gradient $\phi$ of KL divergence gradient, which is proved to be proportional to the kernelized Stein operator within the unit ball of a Reproducing Kernel Hilbert space (RKHS)~\cite{Blundell:2015:WUN:3045118.3045290}. Accordingly, the SVGD update equations are given as,
\begin{equation}
\label{eq:svgd}
\theta_{i}^{t+1} = \theta_{i}^{t} + \epsilon_{t}\phi(\theta_i^{t})
\end{equation}
where
\begin{equation}
\label{eq:KSD}
\phi(\theta) = \frac{1}{n}\sum_{j=1}^{n}\left[k(\theta_j^{t},\theta) \underbrace{\nabla_{\theta_j^{t}} \left(\log p(\theta_{j}^{t}) + \log p(\mathcal{D}, \mathcal{R}|\theta_{j}^{t})\right)}_{\mathrm{gradient}} + \underbrace{\nabla_{\theta_j^{t}}k(\theta_{j}^{t},\theta)}_{\mathrm{repulsive\ force}}\right]
\end{equation}
where $i$ represents particle index, $\epsilon_{t}$ is the step size at $t$ iteration, and $k(x, \cdot)$ represents a positive definite kernel (e.g., radial basis function). As a result, an ensemble of DNNs corresponding to $n$ parameter particles $\{\theta_i\}_{i=1}^n$ are trained by SVGD, where the ``gradient" term moves the particles towards high-density regions of the posterior and the ``repulsive force" term imposes diversity and avoids particle collapsing. Compared to parametric VI methods, the particle-based SVGD is able to capture multi-modal posteriors.

\subsection{Physics-constrained likelihood formulation}
In realistic applications, a model only approximates reality and has model-form errors. Hence, it is natural to formulate the model constraints in a probabilistic way to reflect inadequacy of a model. Similar to the constrained Bayesian approach proposed by Wu et al.~\cite{wu2019adding}, the physical equations $\mathcal{R} = 0$ here are formulated as soft constraints, being a part of the likelihood function. We assume that the residual of governing equations obey a zero-mean Gaussian distribution, 
\begin{equation}
p(\mathcal{R}(\mathbf{u}^\theta, P^\theta)|\theta) \sim \mathcal{N}(\mathbf{0}, \Sigma_R)
\end{equation}
where covariance matrix $\Sigma_R$ is is control parameter reflecting our confidence in the physical model. As the Navier-Stokes equations well describe the fluid dynamics in general, a small variance ($\sigma = 10^{-4}$) is specified in this work. Without loss of generality, the sparse observation data errors can be assumed to follow zero-mean Gaussian distributions. Therefore, the log-likelihood function can be explicitly written as the sum of log data likelihood and log equation likelihood,
\begin{equation}
\label{eq_likelihood}
\log(\mathrm{Likelihood}) = \log p(D|\theta, \Sigma_{D}) + \log p(\mathcal{R}|\theta, \Sigma_{R}),
\end{equation} 
where the data covariance matrix $\Sigma_{D} = \mathrm{diag}(\sigma_D)$ are learnable parameters, which can be learned from the data. The prior distribution of the data variance $\sigma_D$ is modeled as an inverse Gamma distribution and the prior of DNN parameters are assumed as a student's T distribution. The physics-constrained SVGD algorithm can be summarized in Algorithm~\ref{alg3}.\\

\begin{algorithm}[H]
	\label{alg3}
	\small
	\SetAlgoLined
	\KwResult{DNN parameters $\theta$ and learnable data variance $\Sigma_D$.}
	Sample prior distributions for $\theta$ and $\Sigma_D$ with $n$ particles\;
	\For{$i = 1 : n_t$}
	{
		1. Calculate log posterior: $L(\theta) = \log p(\theta)+ \log p(\Sigma_{D}) + \log p(\mathcal{D}|\theta, \Sigma_{D}) + \log p(\mathcal{R}|\theta, \Sigma_{R})$\;	
		2. Calculate the gradient $\nabla_{\theta}$ by back-propagation\;
		3. Choose an appropriate kernel function $k(\theta, \cdot)$, and calculate kernel stein operator $\phi$\;
		4. Update $\theta$ and $\Sigma_D$ by stochastic gradient descent (e.g., Adam)\;
	}
	\caption{Physics-Constrained Stein Variational Gradient Descent}
\end{algorithm}

\subsection{Forward propagation with estimated uncertainty}
After training, the physics-constrained Bayesian DNN can be used to reconstruct the flow field given sparse data and high-resolution coordinates by forward propagation. In the SVGD algorithm, an ensemble of trained DNNs will be obtained from the particle-based posterior approximation. Although the concrete form of the posterior is unknown, the statistics of the flowfield predictions can be estimated by the network ensemble using the Monte Carlo method. For example, the mean velocity field $\overline{\mathbf{u}}^\theta$ is computed by,
\begin{equation}
\label{eq:mean}
\overline{\mathbf{u}}^{\theta}  = \mathbb{E}\left[\mathbf{u}(\mathbf{x}, t)\right] \approx \frac{1}{N} \sum_{i=1}^{n}\mathbf{u}^{\theta_i}(\mathbf{x}, t)
\end{equation}
where $n$ is the number of DNNs indexed by $\theta_{i}$. The variance field (reflecting reconstruction uncertainty) is computed based on the law of total variance as shown in Ref.~\cite{zhu2018bayesian}, where conditional covariance is defined as, 
\begin{equation}
\label{eq:covariance}
\begin{aligned}
    \text{Cov}(\mathbf{u}^{\theta}|\mathbf{x},t,D)  &= \mathbb{E}_{\mathbf{w},\sigma_{D}}\left[\text{Cov}(\mathbf{u}^{\theta}|(\mathbf{x},t;\mathbf{w},\sigma_{D}))\right] + \text{Cov}_{\mathbf{w},\sigma_{D}}(\mathbb{E}\left[\text{Cov}(\mathbf{u}^{\theta}|(\mathbf{x},t;\mathbf{w},\sigma_{D}))\right]\\
    &= \mathbb{E}_{\sigma_{D}}\left[\Sigma_{D}\right] + \mathbb{E}_{\mathbf{w}}\left[\mathbf{u}^{\theta}(\mathbf{x},t)^\top\mathbf{u}^{\theta}(\mathbf{x},t)\right] - \mathbb{E}_{\mathbf{w}}\left[\mathbf{u}^{\theta}(\mathbf{x},t)\right]\mathbb{E}^\top_{\mathbf{w}}\left[\mathbf{u}^{\theta}(\mathbf{x},t)\right]\\
    &\approx \frac{1}{N} \sum_{i=1}^{n}\left(\Sigma_{D_i} + \mathbf{u}^{\theta}_{i}(\mathbf{x},t)\mathbf{u}^{\theta}_{i}(\mathbf{x},t)^\top\right) - \left(\frac{1}{N} \sum_{i=1}^{n}\mathbf{u}^{\theta}_{i}(\mathbf{x},t)\right)\left(\frac{1}{N} \sum_{i=1}^{n}\mathbf{u}^{\theta}_{i}(\mathbf{x},t)\right)^\top
\end{aligned}
\end{equation}
where $\Sigma_D^{i} \sim p(\Sigma_{D}|D)$. With the defined mean and variance, a probabilistic flow reconstruction result can be obtained.

\section{Result}
\label{result}
\subsection{Overview}
Several flow cases with two idealized vascular geometries (i.e., stenosis and aneurysm bifurcation) are investigated to demonstrate the performance of the proposed method for flow reconstruction from sparse data. In this study, data are generated by sampling the fully-resolved CFD solutions on sparse locations. We begin our numerical experiments by reconstructing the flow with noise-free data using deterministic physics-constrained (PC) deep learning (cases 1 \& 2). Then we evaluate our proposed physics-constrained Bayesian neural network (PC-BNN) on the same flow reconstruction problems but with noisy data (cases 3 \& 4). Both the reconstructed mean flow fields and uncertainties for different data noise levels are investigated. 

A fully-connected network structure of 3 layers and 20 neurons per layer is built for all the flow cases. The Swish activation function~\cite{ramachandran2017searching} is specified in each layer except the output one, where a linear activation is applied. For both deterministic and probabilistic formulations, the Adam optimizer is used for training, where the batch size and initial learning rate are set as 50 and $1\times 10^{-3}$, respectively. In the probabilistic formulation, the prior of NN parameters $\theta$ is given by a Student's t-distribution $\theta \sim \mathrm{St}(\theta|\mu, \lambda, \nu)$, where $\mu=0, \lambda = 2a_0, \nu = a_0/b_0$. The shape and rate parameters $a_0$ and $b_0$ are specified as $a_0=1$ and $b_0=0.04$, respectively. Furthermore, data uncertainties (noise) are assumed homoscedastic, and thus the covariance matrix of data likelihood is a diagonal matrix where the prior distribution of the diagonal term is assumed to be an Inverse Gamma IG$(\beta|a_1,b_1)$ with $a_1 = 2$ and $b_1 = 1\times 10^{-6}$. The equation likelihood is assumed to follow a Gaussian distribution with variance $\sigma^2 = 1\times10^{-4}$. To perform SVGD, an ensemble of five NN samples are generated based on the prior. The Bayesian DNN and physics-constrained SVGD are implemented in the PyTorch platform~\cite{paszke2019pytorch}. The training of $6\times10^{4}$ SGD iterations is performed for deterministic cases, while $1.2\times10^{5}$ SGD iterations for probabilistic cases, on an NVIDIA GeForce RTX 2080 Graphics Processing Unit (GPU) card. The code and dataset for this work will become available at {https://github.com/Jianxun-Wang/Physics-constrained-Bayesian-deep-learning} upon publication.

\subsubsection{Case 1: deterministic reconstruction of flow in idealized stenosis}
\label{case1}
\begin{figure}[htp]
	\centering
    \subfloat[CFD benchmark and data locations]{\includegraphics[width =
	0.4\textwidth]{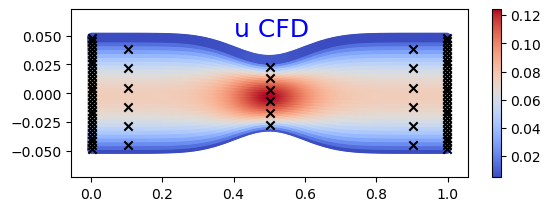}}
	\subfloat[DNN (Navier-Stokes constrained)]{\includegraphics[width =
	0.4\textwidth]{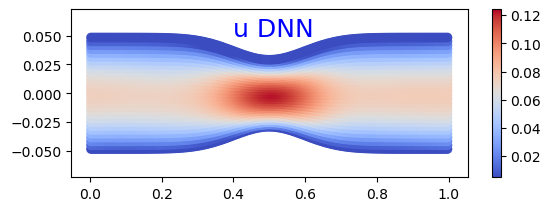}}
	\vfill
	\subfloat[DNN (continuity constrained)]{\includegraphics[width =
	0.4\textwidth]{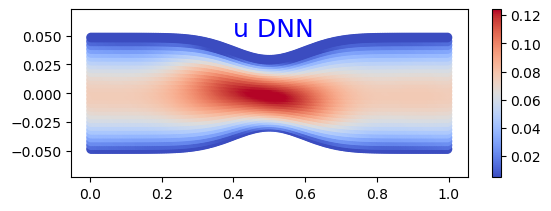}}
	\subfloat[DNN (data only)]{\includegraphics[width =
	0.4\textwidth]{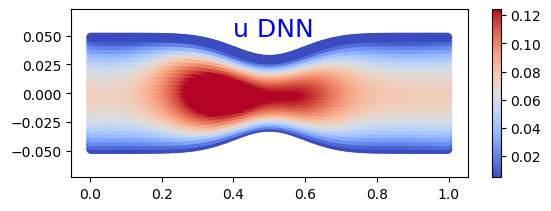}}
	\caption{Comparison of (a) the CFD benchmark (ground truth) with deterministic flow reconstruction results by (b) Navier-Stokes constrained DNN, (c) Divergence-free constrained DNN, and (c) purely data-based DNN for a stenotic flow.}
	\label{fig:stenosis_det}
\end{figure}

In case 1, we aim to reconstruct the flow field in an idealized stenotic vessel from velocity data on very sparse locations (marked as ``x" in Fig.~\ref{fig:stenosis_det}a).  As mentioned above the data are generated directly from the CFD benchmark (Fig.~\ref{fig:stenosis_det}a) without adding any noise. Nonetheless, the data are too sparse to provide sufficient information for flow reconstruction.  
As shown in Fig.~\ref{fig:stenosis_det}d, where the DNN is trained solely based on data, the reconstructed flow is not physical at all and flow features at the tapered region are distorted. If the training process is constrained by the divergence-free condition, i.e., continuity equation, the result can be significantly improved (e.g, flow speed decreasing due to increased radius can be observed in Fig.~\ref{fig:stenosis_det}c), but notable discrepancies still exist compared to the CFD benchmark. When the training is constrained by both continuity and momentum equations (i.e., full Navier-Stokes equations) with boundary conditions, the velocity contour of the reconstructed flow field is almost identical to the CFD benchmark (see Fig.~\ref{fig:stenosis_det}b). The relative reconstruction errors from the purely data-based learning, divergence-free constrained learning, and Navier-Stokes constrained learning are $24.8\%$, $11.3\%$, and $5.6\%$, respectively. The results showed here clearly demonstrate that proper physical constraints can provide additional information to compensate for data insufficiency and enable physical flow reconstruction using limited measurements.

\subsubsection{Case 2: deterministic reconstruction of flow in idealized aneurysm bifurcation}
\label{case2}
To further demonstrate effectiveness of the physical constraints for super-resolution, a more complex flow (i.e., flow in an idealized aneurysm bifurcation) is considered here. The model has a perfect ``T" shape, where the flow starts from the bottom of the vertical tube and goes out through two 90$^{\circ}$ bifurcation arms, driven by a pressure drop $\Delta P = 0.1$. The dome at the end of the input tube represents an idealized terminal aneurysm. 
\floatsetup[figure]{style=plain,subcapbesideposition=top}
\fboxsep= 0pt
\fboxrule=0.1pt
\begin{figure}[htp]
	\centering
	\sidesubfloat[]{\fbox{\includegraphics[width =
	0.3\textwidth]{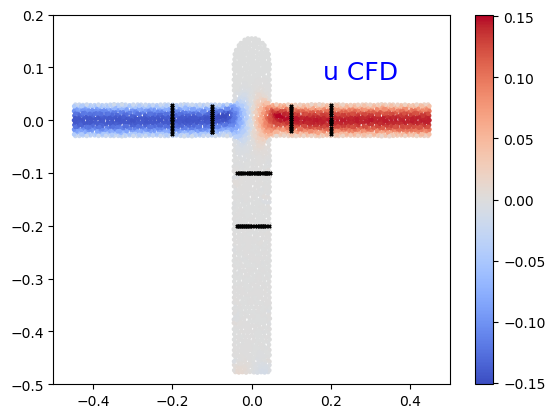}
	\includegraphics[width =
	0.3\textwidth]{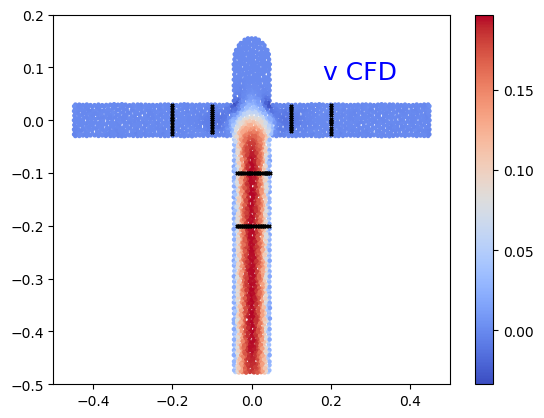}
	\includegraphics[width =
	0.3\textwidth]{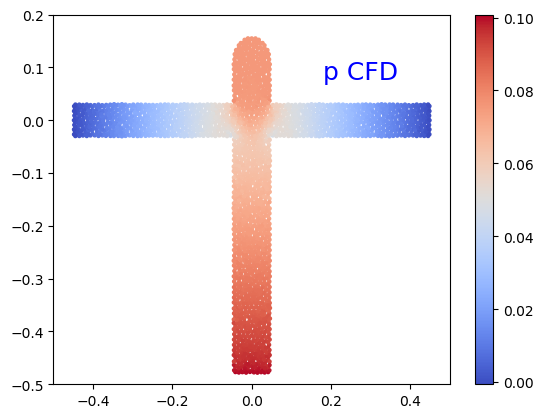}}}
	\par\medskip
	\sidesubfloat[]{\fbox{\includegraphics[width =
	0.3\textwidth]{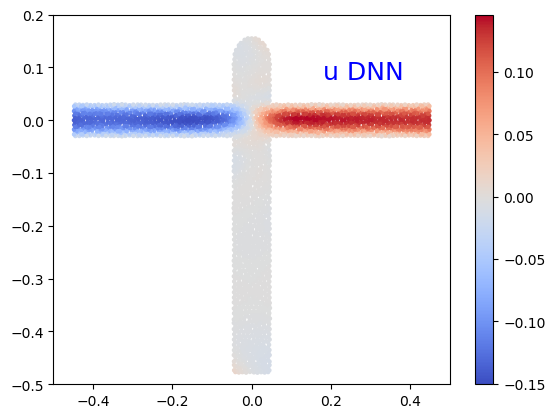}
	\includegraphics[width =
	0.3\textwidth]{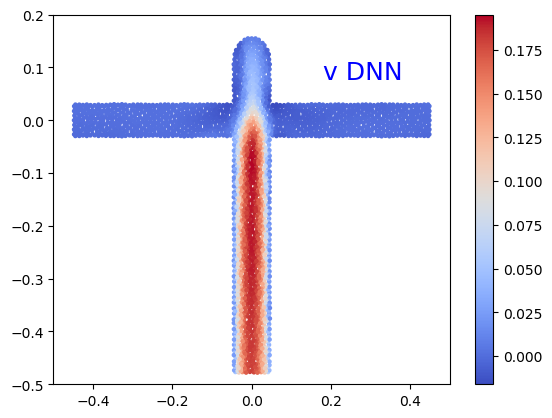}
	\includegraphics[width =
	0.3\textwidth]{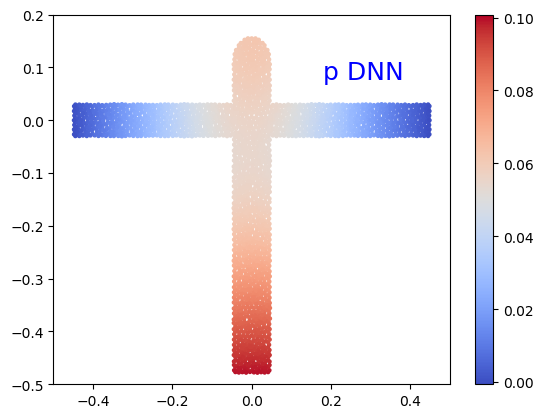}}}
	\par\medskip
	\vfill
	\sidesubfloat[]{\fbox{\includegraphics[width =
	0.3\textwidth]{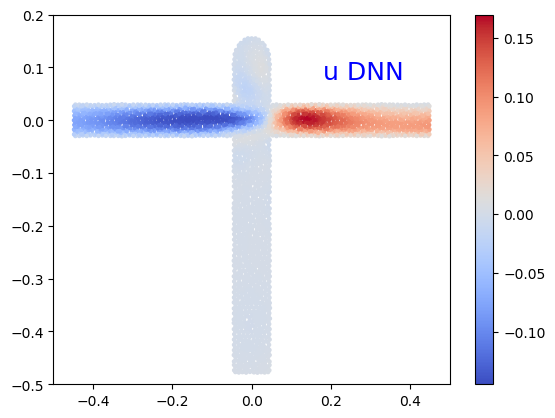}
	\includegraphics[width =
	0.3\textwidth]{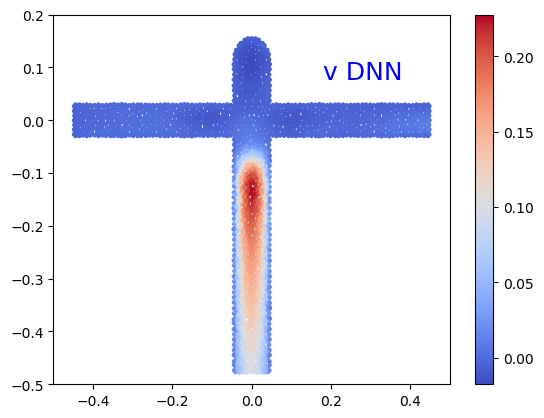}
	\includegraphics[width =
	0.3\textwidth]{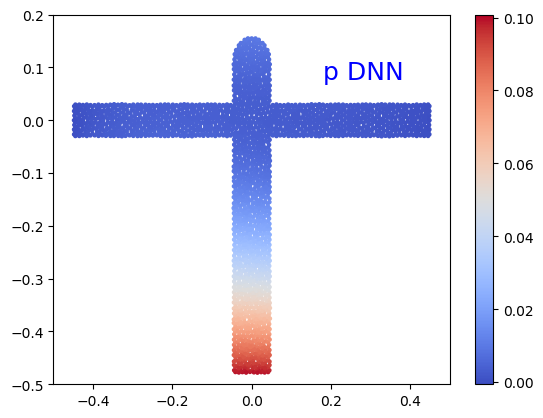}}}
\caption{Comparison of (a) CFD benchmark (with locations of labelled data) with deterministic flow reconstruction by (b) Physics-constrained DNN and (c) purely data-based DNN for an aneurysm bifurcation flow.}
\label{fig:Tjunction_det}
\end{figure}
The data were obtained by probing the CFD velocity field on only six slices, which are very sparse in general (see Fig.~\ref{fig:Tjunction_det}a). Following the physics-constrained learning, where the wall boundary condition is enforced softly, the reconstructed velocity and pressure fields (see Fig.~\ref{fig:Tjunction_det}b) agree well with the CFD benchmark. For the sake of comparison, the purely data-based learning results are also presented in Fig.~\ref{fig:Tjunction_det}c, where the reconstructed velocity fields significantly differ from the CFD benchmarks. It is worthwhile to note that the purely data-based DNN fails to reconstruct the pressure field since no pressure data are used for training. However, the physics-constrained DNN can reasonably capture the general patterns of pressure field because of the constraints on the relation between pressure and velocity, imposed by the Navier-Stokes equations. Quantitatively, the relative reconstruction errors ($\frac{|f_{DNN} - f_{CFD}|}{|f_{CFD}|}$) in $u, v,$ and $P$ from the purely data-based DNN are $35.1\%, 40.5\%$, and $69.9\%$, respective, while for Navier-Stokes constrained learning, the relative errors can be reduced to $13.7\%, 12.1\%$, and $12.8\%$. These comparisons show that the PC-NN remarkably improves the reconstruction accuracy for velocity, and it also can infer the pressure field with the same level of accuracy, where no data are used for training.

\subsubsection{Case 3: Bayesian reconstruction of flow in idealized stenosis} 
\label{case3}
In the two cases presented above, the flow fields are reconstructed from noise-free data based on deterministic physics-constrained learning. However, when the data are not only sparse but also noisy, the uncertainty due to measurement noises should be reflected in the reconstructed flow. Hence, a physics-constrained Bayesian neural network (PC-BNN) is trained on the sparse, noisy data to enable robust flow reconstruction with quantified uncertainties. The same flow in Case 1 is reconstructed but with noisy data sampled from the CFD benchmark solutions that are corrupted by Gaussian noises of different levels. Similarly, the noisy flow is ``observed" only at a few locations indicated by ``x". 
\begin{figure}[htp]
	\centering
	\subfloat[corrupted CFD and data locations]{\includegraphics[width =
	0.4\textwidth]{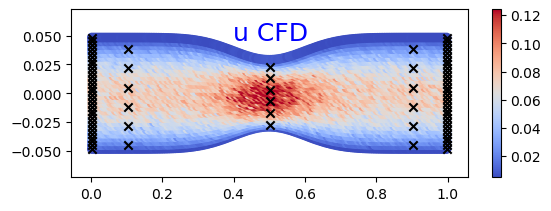}}
	\subfloat[DNN (noisy data only)]{\includegraphics[width =
	0.4\textwidth]{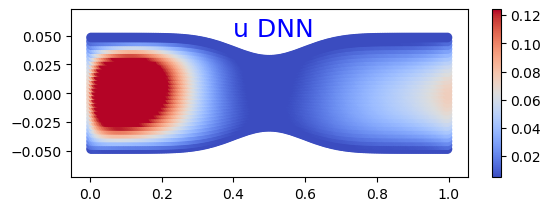}}
	\vfill
	\subfloat[PC-BNN mean]{\includegraphics[width =
	0.4\textwidth]{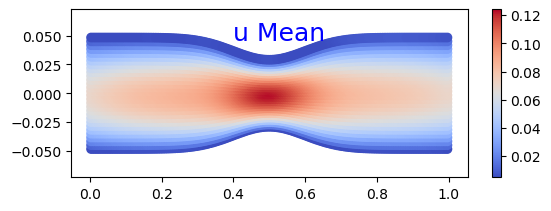}}
	\subfloat[PC-BNN uncertainty]{\includegraphics[width =
	0.4\textwidth]{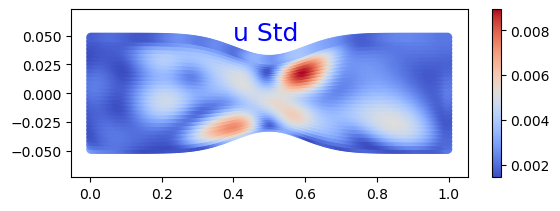}}
		\caption{Comparison of reconstruction results of the stenotic flow with sparse, noisy data ($10\%$ noise) between (b) purely data-based DNN prediction and (c) mean velocity field reconstructed by PC-BNN. The uncertainty of PC-BNN based flow reconstruction (i.e., standard deviation field) is shown in panel (d).}
	\label{fig:stenosis_uq}
\end{figure}
Figure~\ref{fig:stenosis_uq} shows the flow reconstruction results by PC-BNN, while the purely data-based solution is also plotted for comparison. We can see that the flow field corrupted by $10\%$ Gaussian noise becomes unsmooth (Fig.~\ref{fig:stenosis_uq}a), and the purely data-based flow estimation (Fig.~\ref{fig:stenosis_uq}b) fails to capture any physical flow patterns. The relative reconstruction error increases to $83.3\%$ (Fig.~\ref{fig:stenosis_det}d). This is expected since the data are lack of both quantity and quality. In contrast, the mean-field of the reconstructed flow by PC-BNN (Fig.~\ref{fig:stenosis_uq}c) is in a good agreement with the CFD benchmark (Fig.~\ref{fig:stenosis_det}a) and the noise can be notably reduced as well. The relative error of the mean reconstructed field is reduced from $83.3\%$ to $6.9\%$ by introducing the Navier-Stokes equation constraint. Moreover, the uncertainty of the reconstructed flow can be reasonably estimated as shown by the standard deviation (std) field in Fig.~\ref{fig:stenosis_uq}d. We have studied the reconstruction performance given different data noise levels ($5\%, 10\%, 15\%$), and the prediction results and uncertainties are summarized in Table~\ref{tab:apptab1} in Appendix. The reconstruction error of the purely data-based DNN remarkably increases with increased data noise. Although the accuracy of the PC-BNN predictions also slightly decreases with the increased noise level, the performance is still satisfactory and flow physics can be captured reasonably well. It is important to note that the mean and maximum std of the reconstructed field also increases as the data noise becomes larger, demonstrating that the uncertainty of the reconstructed results can be well estimated by the PC-BNN, which reflects the effect of data noises.

\subsubsection{Case 4: Bayesian reconstruction of flow in idealized aneurysm bifurcation}
\begin{figure}[htp]
	\centering
	\subfloat[Corrupted CFD and data locations]{\includegraphics[width =
	0.38\textwidth]{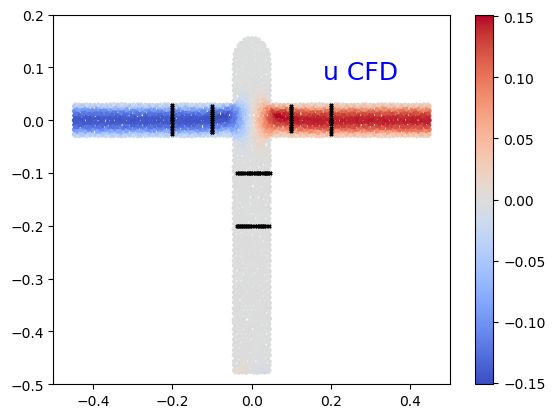}}
	\subfloat[DNN (noisy data only)]{\includegraphics[width =
	0.4\textwidth]{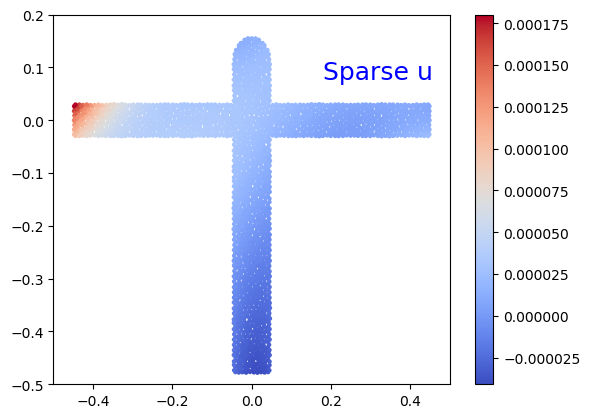}}
	\vfill
	\subfloat[PC-BNN mean]{\includegraphics[width =
	0.38\textwidth]{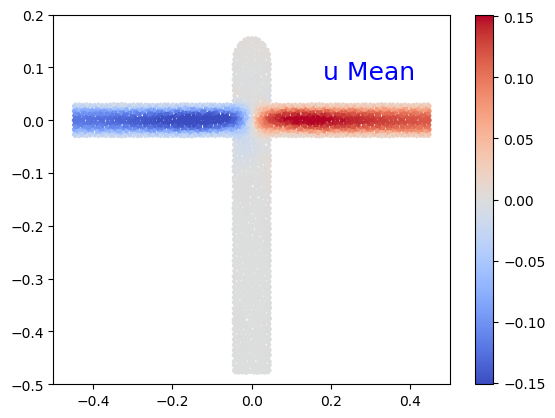}}
	\subfloat[PC-BNN uncertainty]{\includegraphics[width =
	0.38\textwidth]{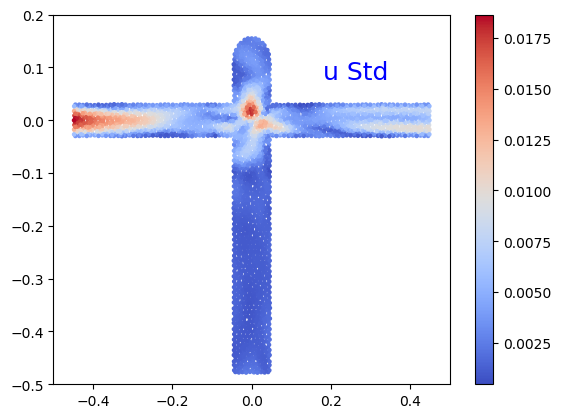}}
		\caption{Comparison of reconstruction results of the bifurcation flow with sparse, noisy data ($10\%$ noise) between (b) purely data-based DNN prediction and (c) mean velocity field reconstructed by PC-BNN. The uncertainty of PC-BNN based flow reconstruction (i.e., standard deviation field) is shown in panel (d).}
	\label{fig:junction_uq_noise10}
\end{figure}

Lastly, the PC-BNN is applied to reconstruct the bifurcation flow from noisy, sparse data. Gaussian noises with different variances are added onto the CFD solution and six sections of the corrupted flow data (see Fig.~\ref{fig:junction_uq_noise10}) are used for training. 
\label{case4}
Figure~\ref{fig:junction_uq_noise10}a shows the corrupted CFD solution and marks the locations of training data by ``x". For the purely data-based learning, the reconstruction result by noisy data is much worse than that using noise-free data, which is expected (see Fig.~\ref{fig:junction_uq_noise10}b). In contrast, the PC-BNN still accurately captures the flow field and the mean velocity contour shown in Fig.~\ref{fig:junction_uq_noise10}c agrees with the CFD benchmark in Fig.~\ref{fig:Tjunction_det}a. Furthermore, the reconstruction uncertainty introduced by data noise can be reflected by the std field in Fig.~\ref{fig:junction_uq_noise10}d. The uncertainty is large at the left outlet region, indicating that the prediction at this area has low fidelity. Similarly, the performance of the PC-BNN on the data with different noise levels are studied, and the reconstruction accuracy and uncertainty are summarized in Table~\ref{tab:apptab1}. The same trend as shown in Case 3 can be found: the mean reconstruction error and uncertainty will increase as the noise grows. The results from both Case 3 \& 4 show that the proposed PC-BNN can accurately reconstruct a high-resolution flow field from sparse and noisy data, and the prediction uncertainty can also be estimated.

\section{Discussion on the role of data and constraints in deep learning}
\label{discussion}
In this work, the numerical results have demonstrated that a high-resolution flow field can be recovered following physics-constrained learning with sparse data and known physical constraints (i.e., Navier-Stokes equations). However, a previous work~\cite{SUN2019112732} in the context of surrogate modeling has shown that flow solutions of the Navier-stokes equations can be obtained from physics-constrained deep learning even without any labeled data if the boundary conditions are imposed properly. Therefore, it is interesting to know what benefit can be gained by introducing additional sparse labeled data into PDE-constrained learning. Taking the stenotic flow as an example, we conducted a comparison study between the \emph{data-free} PDE-constrained learning and \emph{weakly data-based} (i.e., sparse data-based) PDE-constrained learning, where boundary conditions are both imposed softly with a penalty parameter $\lambda = 0.1$,
\begin{figure}[htp]
	\centering
	\includegraphics[width =
	0.32\textwidth]{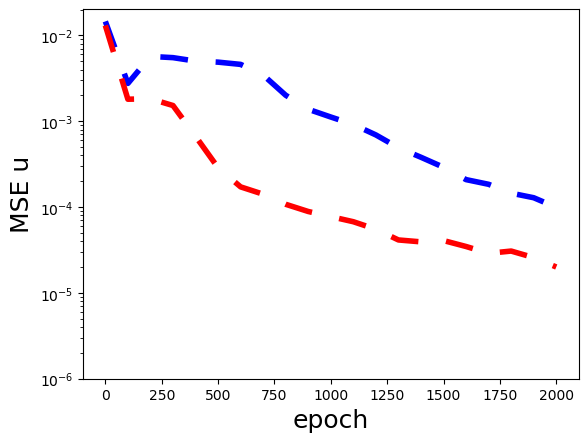}
	\includegraphics[width =
	0.32\textwidth]{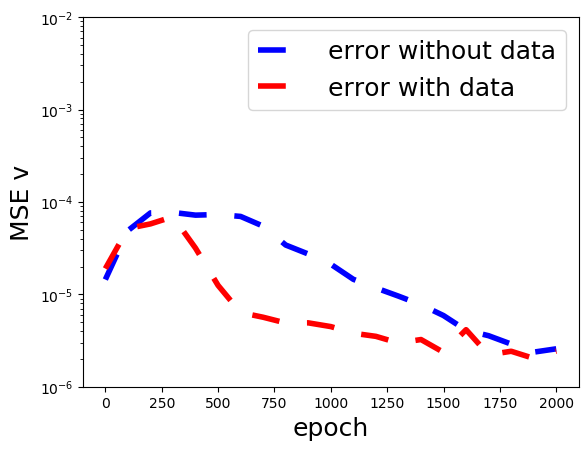}
	\includegraphics[width =
	0.32\textwidth]{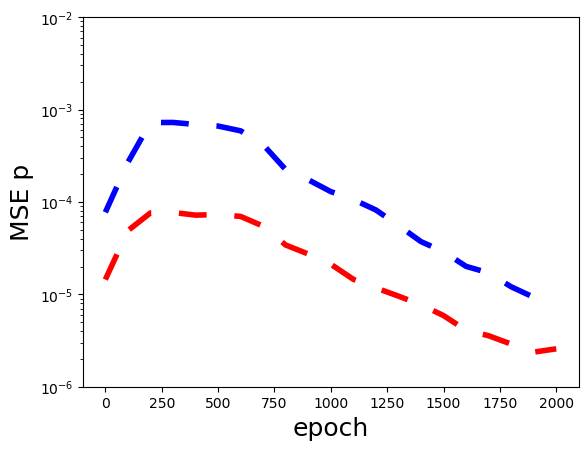}
	\caption{Test error histories of the physics-constrained learning with and without using training data. The mean square errors (MSE) of $u$ (left), $v$ (middel), and $P$ (right) predictions are compared.}
	\label{fig:test_error}
\end{figure}
Figure~\ref{fig:test_error} shows the histories of test errors for velocity and pressure versus the number of training epochs. The test error of the weakly data-based learning (red dashed line) decreases much faster than that of the data-free learning (blue solid line). With the same number of training epochs, the prediction error from sparse data-based, physics-constrained learning is about one order of magnitude lower than purely physics-constrained learning without any labeled data. The comparison indicates that adding some labeled data would further improve the equation-constrained learning, which is consistent with the intuition since more information is used for the neural network training.

\section{Conclusion}
\label{conclusion}
The objective of this work is to reconstruct a high-resolution flow field from sparse and possibly noisy data. To achieve this goal, we proposed a physics-constrained Bayesian deep learning framework, where the likelihood function is constructed based on the measurement uncertainty and model inadequacy. Stein variation gradient descent is used to enable efficient Bayesian learning. The proposed approach is able to reconstruct the flow field with estimated uncertainties particularly when data are corrupted with measurement noise. Numerical experiments were conducted on a number of flow reconstruction cases with idealized vascular geometries, where synthetic data are used to evaluate the performance of the proposed method. We have demonstrated that the constraints of a physical model can significantly improve the reconstruction results from limited clean data. When the data are noisy, our proposed PC-BNN can accurately predict the mean flow field, meanwhile reasonably estimate the prediction uncertainties corresponding to different data noise levels.          


\section*{Acknowledgement}
LS and JXW gratefully acknowledge support from the National Science Foundation (NSF contract CMMI-1934300) and Defense Advanced Research Projects Agency (DARPA) under the Physics of Artificial Intelligence (PAI) program (contract HR00111890034). LS would also acknowledge partial funding support by graduate fellowship from China Scholarship Council (CSC) in this effort. The authors would like to thank Dr. Nicholas Zabaras and Dr. Yinhao Zhu for their helpful discussions during this work.

\section*{Appendix}
\label{Apendix}
The performance of PC-BNN with data of different noise levels are summarized in Table~\ref{tab:apptab1}. Both the test errors of the mean fields and reconstruction uncertainties are listed. The uncertainty is given by std norm, where both the mean and max values are presented.
\begin{table}[H]
\begin{center}
\begin{tabular}{| c | c | c | c | }
  \hline
  \multicolumn{4}{|c|}{\textbf{Reconstructed $u$ velocity field of stenotic flow}}\\
  \hline 
  \multicolumn{2}{|c|}{level of noise} & relative mean error & uncertainty (mean/max std) \\
  \hline
  \multirow{2}*{5\% noise}      & data only & $0.623$& not applicable \\\cline{2-4}
       & PC-BNN &$0.041$ &$0.038$/$0.076$ \\
  \hline 
   \multirow{2}*{10\% noise}      & data only & $0.833$& not applicable \\\cline{2-4}
       & PC-BNN &$0.069$ &$0.050$/$0.083$ \\
  \hline 
   \multirow{2}*{15\% noise}      & data only & $0.941$& not applicable \\\cline{2-4}
       & PC-BNN &$0.100$& $0.064$/$0.125$
\\
  \hline 
  \multicolumn{4}{|c|}{\textbf{Reconstructed $u$ velocity field of bifurcation flow}}\\
  \hline 
  \multicolumn{2}{|c|}{level of noise} & relative mean error & uncertainty (mean/max std)\\
  \hline
  \multirow{2}*{5\% noise}      & data only & $1.000$& not applicable   \\\cline{2-4}
       & PC-BNN &$0.132$ & $0.046$/$0.210$   \\
  \hline 
   \multirow{2}*{10\% noise}      & data only & $1.000$& not applicable  \\\cline{2-4}
       & PC-BNN & $0.126$ & $0.062$/$0.253$\\
  \hline 
  \multirow{2}*{15\% noise}      & data only & $1.000$& not applicable   \\\cline{2-4}
       & PC-BNN & $0.173$ &$0.067$/$0.325$\\
  \hline 
\end{tabular}
\end{center}
\caption{Mean errors and uncertainties of reconstructed $u$ velocity fields from sparse, noisy data using physics-constrained Bayesian neural network (PC-BNN). Note that both the error and std are normalized by the corresponding CFD benchmark solution.}
\label{tab:apptab1}
\end{table}
\end{document}